# Chemical reaction mechanism of pre-curing process of two-component adhesive based on deformation behavior for automobile hood

Jia Li [a, *], Jiao Li [a], Li Huang [b], Feng Gao [c], Chao Peng [d]

a Chongqing Industrial Polytechnic College, China

b China Automotive Engineering Research Institute Co., Ltd.

c Chongqing Chang'an Automobile Co., Ltd., China

d Chongqing Chassis Systems Branch of China Chang'an Automobile Group Co., Ltd

**Abstract:** Simulating the actual conditions of assembling and baking-curing for the automobile hood in the paint shop, the shearing test is carried out on the joint which bonded under different pre-curing processes with two-component adhesives of acrylic and epoxy resin respectively. The pre-curing strength is obtained, which used to analyze the relationship between the pre curing strength and time. The automobile hoods with different pre-curing strength are baking with high temperature in the paint shop. After that, the deformation of different areas of the hood is measured with the help of the gauge, and the deformation characteristics of the hood after baking are acquired with comparative analysis. Combining the components of the two-component adhesives and the DSC test, the pre-curing mechanism of different adhesive systems are studied. Therefore, the key factors and regular pattern of the deformation control for the automobile hood are obtained, which is influenced by the pre-curing processes of the two-component adhesive. The results indicate that, the finally deformation of the hood after high-temperature baking varies with the pre-curing time. The key of the pre-curing time of acrylic adhesives at the room temperature lies in the chain initiation stage of the free radical polymerization reaction. Due to the influence of the chemical properties of methyl acrylate and its initiator, the pre-curing reaction process induced by free radical polymerization is very fast. The shear strength of this joint can reach to 3.67 MPa with a pre-curing time of 1 h, which quickly achieving the pre-cure strength required for deformation control. For epoxy resin adhesives, the rate-determining step in the pre-curing reaction process is the esterification reaction. Due to the influence of the structure of carboxylic acid, the esterification reaction process is relatively long. The shear strength of this joint can only reach 0.87 MPa after pre-curing for 4 h without external heating. Since the inability to quickly achieve effective pre-curing strength, it is recommended to conduct heating pre-curing treatment at 80 °C to accelerate the pre-curing reaction, which can reduce the occurrence of deformation.

## 1 Introduction

Adhesive bonding is an advantageous connection technology to firmly join similar or dissimilar materials together relying on the bonding force generated by the curing reaction of adhesives, which be widely used in automotive manufacturing [1,2]. During the curing process of the adhesive, the chemical reactions within high-temperature baking may result in the accumulation of internal stresses or dimensional changes at the bonding interface, which leading to contraction and expansion deformation or stress generation in the adhesive joints [3]. Not only does it affect the bonding performance, but it also seriously affects the dimensional accuracy of the adhesive joint.

For automotive manufacturing, it is necessary to consider the strength of the vehicle body on the one hand, it is also need to consider the dimensional accuracy requirements of the exterior surface, especially exterior components such as the automobile hood. The inner and outer panels of the hood are bonded together with adhesive. After the hood is assembled onto the vehicle body, it needs to fit with the front headlights, fenders, and other areas of the car. If there are significant deformation, it will result in noticeable gaps and height differences at the mating positions between two parts, which severely affecting the overall refined appearance of the vehicle [4]. Therefore, there is a high demand for dimensional accuracy after the bonding of the hood.

Different adhesives exhibit different deformation characteristics during the curing process, and the appropriate adhesive can help reduce the amount of deformation. The automotive manufacturers that dominated, such as Tesla, Ford, Volkswagen, etc., choose two-component adhesives curing at room temperature for the bonding of automobile hood [5-7]. Once the A and B components of the adhesive are mixed, the curing reaction begins and quickly forms a certain bonding strength at room temperature. However, complete curing occurs only after high-temperature baking, which resulting in a strong joint.

It is not sufficient to only consider the curing reaction during the baking process at high temperature for the adhesive bonding of automotive hood. In actual production processes, there are several flow operations of the adhesive joint before the high-temperature baking, such as transportation, assembly and immersion in paint shop. At this stage, due to the low friction between the inner and outer panels of the front hood at the bonding locations and the extremely thin thickness (0.7mm) of the outer panel, these areas shall subject to external forces during these flow operations. It will generate additional irregular distortion and deformation to a certain extent. After high-temperature baking, the deformation of these parts becomes particularly prominent under the further expansion and contraction of the metal. This will seriously affect the overall exquisite appearance of the vehicle. The stage before high-temperature baking and complete curing, where the A and B components are mixed at room temperature, is commonly referred to as pre-curing.

Numerous studies are available that investigate the effects of curing temperature and curing time on the mechanical properties of both bulk adhesive materials and adhesively bonded joints [8-10]. Omar Moussa et al as representativity focus on early-age the tensile properties of structural epoxy adhesives subjected to low-temperature curing, and isothermal curing was carried out, with curing temperatures set at 5 ℃, 10 ℃, 25 ℃, 40 ℃, and 70 ℃, and curing times selected between 0.17h and 720h [11].

However, none of these previous works focused on the relationship between the early-age mechanical properties and pre-curing mechanism under low curing temperatures, particularly for the impact on the deformation of adhesive joints in automotive engineering applications.

In this study, the baking deformation of two-component adhesives under different pre-curing processes is investigated by comparison experiments, which based on the production process of adhesive for automotive hoods. The key factors and regular patterns of the deformation control influenced by the curing reaction of two-component adhesives on the hood are analyzed. Differential scanning calorimetry (DSC) is used to investigate the pre-curing reaction process of different adhesive systems, which providing a theoretical basis for optimizing adhesive composition and pre-curing processes.

## 2 Experimental procedures

## 2.1. Materials

The material of the test samples used for the inner and outer panels of the hood is DC01 cold-rolled steel plate with a thickness of 0.7 mm. The test piece is coated with standard anti-rust oil (Beishan Chemical R3304-MJ) which is consistent with the actual vehicle working conditions.

The two-component adhesives used in this experiment are 3M SA9816 based on epoxy and LORD Versilok265/254 based on acrylic. The composition is shown in Table 1. The matrix material of 3M adhesive is epoxy resin, which of LORD is acrylic acid. The mixing ratio of AB components is 4:1.

Table 1 Comparison of the main components of two-component hemming adhesive

| Type | No. | Component A | wt.% | Component B | wt.% |
|---|---|---|---|---|---|
| 3M SA9816 | 1 | (3-Glycidyloxypropyl) triethoxysilane | 1 - 5 | Fatty acids, C18-unsatd., dimers, polymers with 3,3-oxybis(2,1-ethanediyloxy) bis1-propanamine | 30 - 60 |
| | 2 | Quartz (SiO2) | 1 - 5 | Quartz (SiO2) | 10 - 15 |
| | 3 | Resin epoxy | 30 - 60 | Tris(dimethylaminomethyl)phenol | 10 - 15 |
| | 4 | Fiber Glass Wool | 10 - 30 | [Oxybis(2,1-ethanediyloxy)] bis-1-propanamine | 7 - 13 |
| | 5 | 1,4-Bis((2,3-epoxypropoxy)methyl)cyclohexane | 7 - 13 | bis[(dimethylamino)methyl] phenol | 1 - 5 |
| | 6 | Siloxanes and Silicones, di-Me, reaction products with silica | 1 - 5 | Calcium nitrate tetrahydrate | 1 - 5 |
| | 7 | Carbon Black | 0.1 - 1 | Siloxanes and Silicones, di-Me, reaction products with silica | 1 - 5 |
| | 8 | filler | other | Fiber Glass Wool | 1 - 5 |
| | 9 | | | Toluene | < 0.5 |
| | 10 | | | filler | other |
| LORD Versilok 265/254 | 1 | MMA monomer | <30.0 % | Resin epoxy | <55.0 % |
| | 2 | Ethoxylated bisphenol A dimethacrylate | <5.0 % | Benzoyl peroxide | < 5.0 % |
| | 3 | 2-Hydroxyethyl methacrylate phosphate | <5.0 % | Tri-o-cresyl phosphate | < 5.0 % |
| | 4 | titanium dioxide (TiO2) | <5.0 % | filler | other |
| | 5 | Amine drugs | <5.0 % | | |
| | 6 | filler | other | | |

## 2.2. Shearing Test

During the experiment, the conditions of adhesive bonding for the samples are the room temperature of (23 ± 2) °C, relative humidity of (50 ± 5) %. The parameters of curing are at 170 °C± 2 °C for 20 minutes placed in the baking oven. The specimens for mechanical properties testing shall be made according to the provisions of GB/T 7124-2008. In the test joints, the effective bonding length is 12.5 mm, and the thickness of adhesive layer is 0.2 mm~0.3 mm, as shown in Figure 1. After baking and curing, the samples are cooled for 24 hours at room temperature, and then the shear test of the joints are conducted at room temperature with a tensile speed of 10 mm/min.

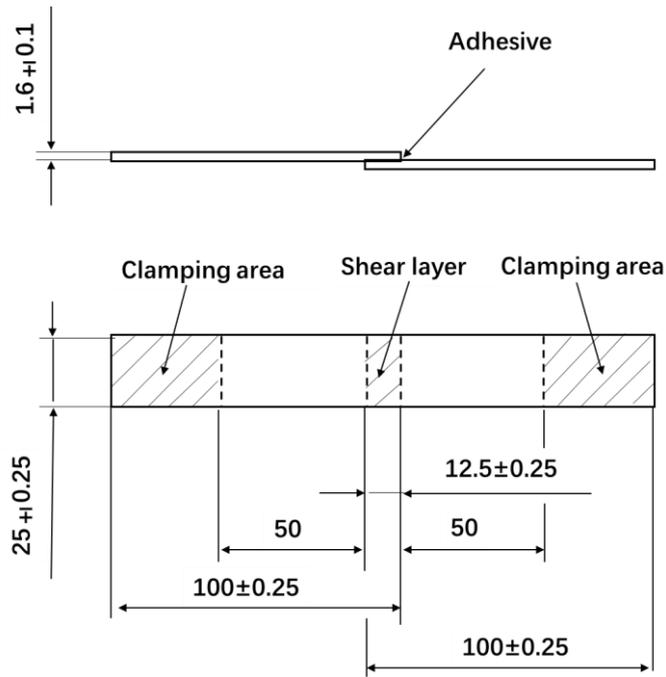

Fig.1 Schematic diagram of the shear strength test of adhesive joint

**2.3 Deformation measurement**

To simulate the real situation, the hoods are immediately transported to the body shop for normal assembly after adhesive bonding. After assembly, the car body will be transported to the pain shop for electrophoresis, baking and curing. Among them, the baking temperature is 170 °C~200 °C, and the baking time is 20 min. Finally, the hoods are removed, and then the deformation are measured on the gauges, with the measurement points located as shown in Figure 2.

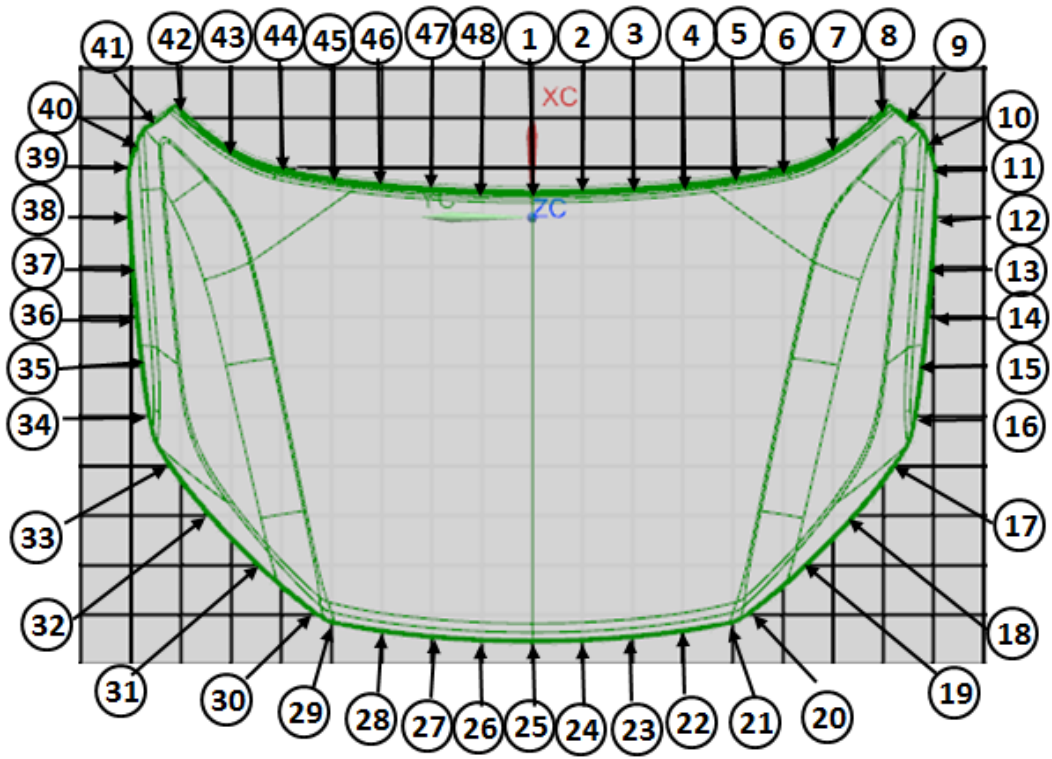

**Fig.2** Schematic diagram of deformation measuring points of the hood

## 3 Result

### 3.1 Performance of adhesive joint under different pre curing time

The shear strength tests are conducted on the adhesive joints which are bonded at different pre-curing times and without high-temperature baking. The test results，namely the pre-curing strength, are shown in Figure 3. The results show that the shear strength of the adhesive joint bonded with LORD is 3.67 MPa, when the pre-curing time is 1 h at room temperature. Until the pre-curing time is extended to 2 h, the adhesive continues to cure, which promotes the pre-curing strength to 7.39 MPa. However, the joints bonded with 3M has almost no strength, even if the pre-curing time reaches to 2 h. When the joints continue to cure for 4 h, the pre-curing strength only increases to 0.87 MPa.

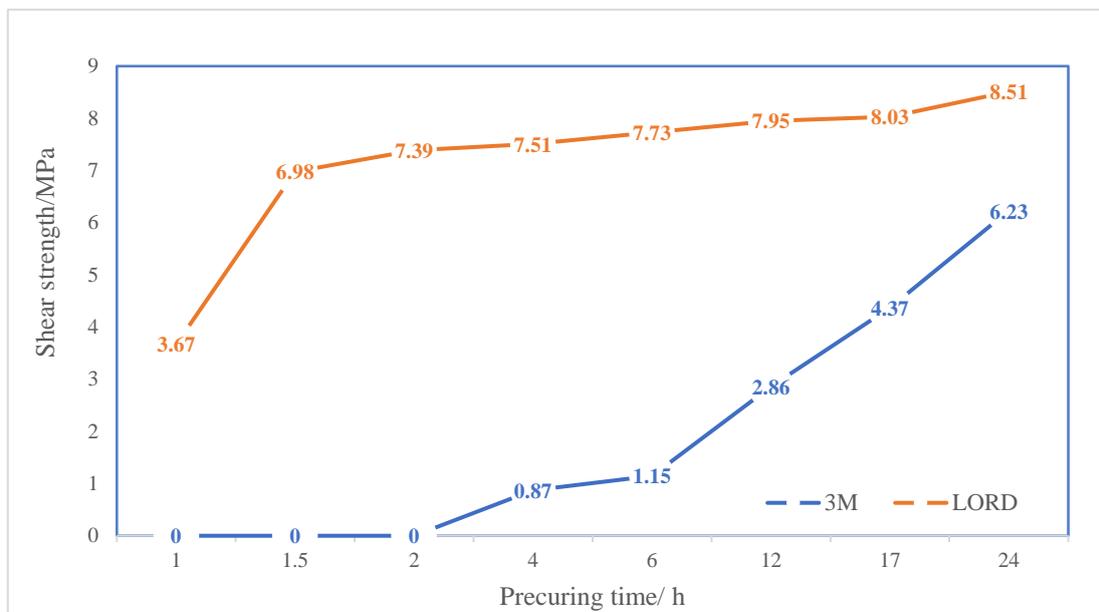

**Fig.3** Shear strength of adhesive joint with different pre-curing time

The production process of the hood of an actual automobile in China's Changan automobile company is investigated. The survey results are shown in Table 2. The results indicate that it takes about 2~3 hours from bonding to transportation to paint shop for baking.

**Table 2** The time from bonding to the painting of the hood

| Number | Time from gluing to bonding/min | Time from bonding to loading in car /min | Time from bonding to paint shop /h |
|---|---|---|---|
| 1 | 17 | 34 | 2.5 |
| 2 | 14 | 26 | 2.4 |
| 3 | 10 | 49 | 2.8 |
| 4 | 7 | 59 | 3.1 |
| 5 | 11 | 47 | 2.75 |

For most enterprises, they generally hope that production efficiency is as fast as possible. Therefore, the pre-curing time is based on 2 h when analyzing the pre-curing strength and front cover deformation in the following.

**3.2 Comparison of the hood deformation**

The hoods are bonded with 3M and LORD adhesive separately, and be assembled onto the vehicle body. After 2 h pre-curing, the hoods are baked in an oven in the paint shop. Then, the curing of the hoods is completed. After that, the hoods are disassembled from the vehicle body and the deformation at each measuring point is measured on the dedicated gauges. The measurement results are shown in Figure 4.

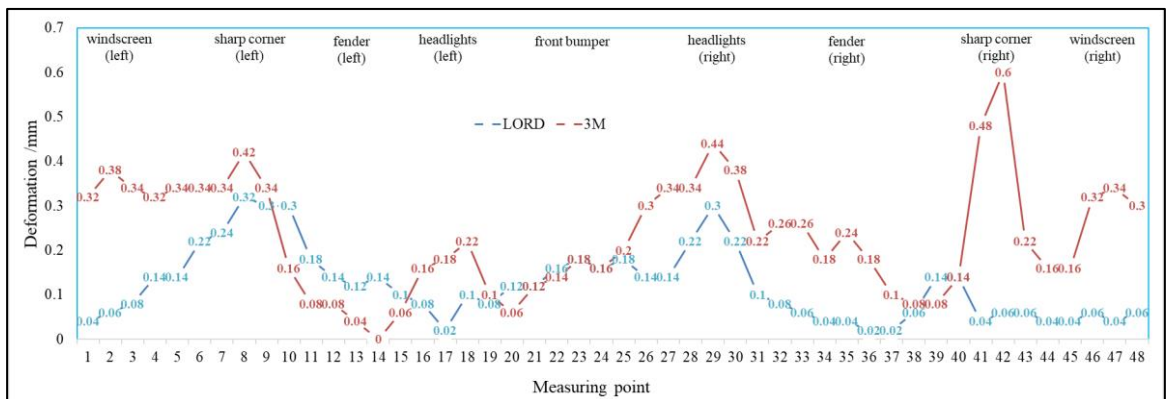

**Fig.4** Deformation measurement results of each area of the hood

It finds that the maximum deformation of the front cover bonded with 3M after baking is reached to 0.6 mm at measuring point 42. According to the body structure, this measuring point is located near the sharp corner of the front windshield area on the right side of the body. The deformation of point 8 at the left sharp corner corresponding to the point 42 of right side was reached to 0.4 mm. In addition, the deformation at point 29, namely the right headlight, is also significant, which is reached to 0.45 mm.

In contrast, the hood bonded with LORD after be baked showed the largest deformation at measuring points 8 and 29, with a deformation of 0.3 mm.

The results indicate that the large deformation areas of the two sets of hoods are concentrated in the Sharp corner areas on both sides of the headlights and front windshield. The main reason for this situation is that the sharp corners of the outer plate of the hood have a greater bending angle, so due to the influence of the stamping process, the spring-back at these positions is more obvious. Furthermore, the inner and outer panels of the front cover will also experience certain warping when heated during the baking process. In addition, this position is usually the clamping part of the front cover handling, and the probability of sheet metal misalignment or warping due to external forces is higher.

In order to further analyze the effect of pre-curing strength on the deformation, the maximum and average deformation of the hoods assembled on real cars bonded with 3M and LORD at different pre-curing times is measured. The measurement results are shown in Table 3.

Table 3 Maximum and average deformation of the hood bonded with different two-component hemming adhesive

| Adhesive | Sample number | Pre-curing time /h | Maximum deformation (height difference) /mm | Average deformation (height difference) /mm |
| --- | --- | --- | --- | --- |
| LORD | 1 | 2.4 | 0.4 | 0.07 |
| | 2 | 2.5 | 0.3 | 0.22 |
| | 3 | 2.7 | 0.4 | 0.13 |
| | 4 | 2.8 | 0.3 | 0.1 |
| | 5 | 3.1 | 0.3 | 0.09 |
| 3M | 1 | 2.5 | 1.5 | 0.39 |
| | 2 | 4.0 | 1 | 0.27 |
| | 3 | 4.1 | 0.8 | 0.19 |
| | 4 | 4.2 | 0.6 | 0.21 |
| | 5 | 4.5 | 0.3 | 0.12 |

The results indicate that the maximum deformation remains at the level of 0.3~0.4mm when the pre-curing time of the hoods bonded with LORD is more than 2 h. By contrast, the maximum deformation decreases with the extension of pre-curing time for the hoods with 3M adhesive, and the maximum deformation can be controlled to a level of 0.3mm until the pre-curing time was extended to 4.5 h.

Due to the influence of the yield behavior and thermal expansion property of metals, slight deformation is inevitable during the production of the hoods. Therefore, the above results indicated that the deformation of the hood in subsequent baking process was relatively small and tended to stabilize at around 0.3 mm, after the pre-curing strength of the joints reached a certain level. Therefore, it can be inferred that the pre-curing strength must reach at least 1 MPa to control the deformation according to shear strength in Figure 3.

## 4 Discussion

According to the above results, it can be seen that the pre-curing strength before baking is a key influencing factor on the adhesive deformation, and the shorter the time it takes to form a certain strength, the easier it is to control the amount of deformation.

### 4.1 Pre-curing reaction process of acrylic adhesive

The matrix of LORD two component adhesive is methyl methacrylate $CH_2=C(CH_3)COOCH_3$. There are double bonds at the chain end of the acrylates, which are prone to polymerization with the free radicals of peroxides. It cures by redox-initiated polymerization of (meth) acrylate monomers [12].

Researchers such as McGinniss V have pointed out that the essence of the curing for acrylate adhesive is free radical chain reaction between acrylic ester and other additive solutions containing double bonds, which going through the stages of chain initiation, chain growth, chain transfer, and chain termination [13].

According to the molecular formula $CH_2=C(CH_3)COOCH_3$ of methyl methacrylate, it belongs to the alkane molecule, and its double bond of carbon (C=C) consists of a σ bond and a π bond [14]. The σ bond has a high degree of overlap, with a cylindrical symmetric distribution along the bond axis. The electron density is concentrated between the two atoms, resulting in a stronger bond and lower polarity. The π bond has a lower degree of overlap, with a symmetric plane along the bond axis. The electron density is more dispersed, distributed above and below the molecular plane. As a result, the π electrons are less tightly bound to the nuclei, allowing for greater electron mobility and a weaker bond. Therefore, the π bond is more easily broken, making it susceptible to polarization by electron-deficient electrophiles.

According to the adhesive composition in Table 1, the initiator that is compatible with acrylic acid is benzoyl peroxide. The chain initiation reaction between methyl methacrylate and benzoyl peroxide initiator is described in Figure 5.

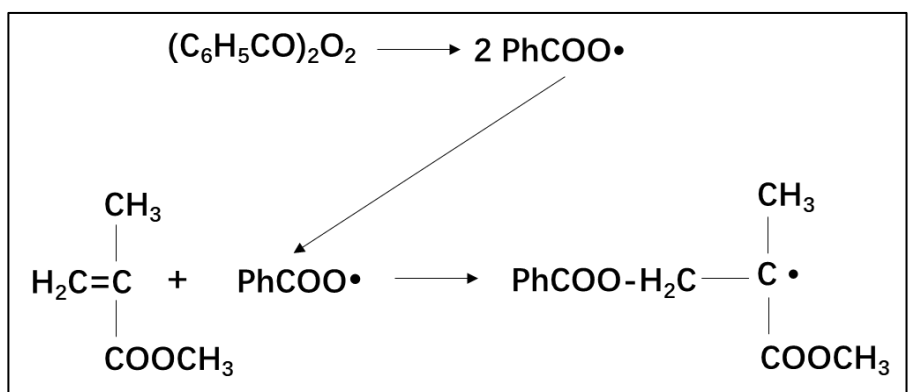

**Fig.5** The chain initiation reaction between methyl methacrylate and benzoyl peroxide initiator

In the B component of LORD acrylic adhesive, the oxygen-oxygen (O-O) bond in the benzoyl peroxide (BPO) molecule is particularly unstable due to the small radius of the (O-O) bond and the close proximity of the lone pair electrons on oxygen, which resulting in significant repulsion. Under the conditions of light and room temperature, the chemical bond can easily break, generating free radicals with unpaired electrons. Free radicals typically have a short lifetime and are reactive intermediates, making them electron-deficient electrophiles. Therefore, the BPO initiator in the B component acts as an electron-deficient electrophile, generating free radicals with unpaired electrons, which can readily react with the acrylic acid reagent in the A component. Thus, the key to the room temperature pre-curing speed of the two-component acrylic adhesive lies in the chain initiation stage.

PhCOO• represents the radicals generated by benzoyl peroxide initiator. The radicals generated by benzoyl peroxide initiator react with the double bonds in methyl methacrylate molecules, which forming new methyl methacrylate radicals. These radicals react with other methyl methacrylate molecules, and gradually extending the length of the polymer chain, ultimately forming a polymer. Therefore, the shear strength of the joints bonded with LORD acrylic adhesive can quickly reach 3.67 MPa at room temperature after pre-curing for 1 h.

**4.2 Pre-curing reaction process of epoxy adhesive**

The main components of 3M two component adhesive are epoxy and compounds such as epoxyethane which with epoxy groups. The typical structure is a saturated heterocyclic compound with a functional group structure of -CH(O)CH-.

Firstly, the curing agent in component B, 3.3'-[oxybis(2,1-ethanediyloxy)] dipropyl amine, acts as a promoter and reacts with the unsaturated fatty acid in component B to form carboxylate anions. The reaction equation is described in Figure 6. Since this reaction is an $S_N2$ nucleophilic substitution reaction, the cleavage of the C-O bond and the nucleophile occur almost simultaneously, resulting in a fast reaction rate that does not significantly affect the overall crosslinking and curing time.

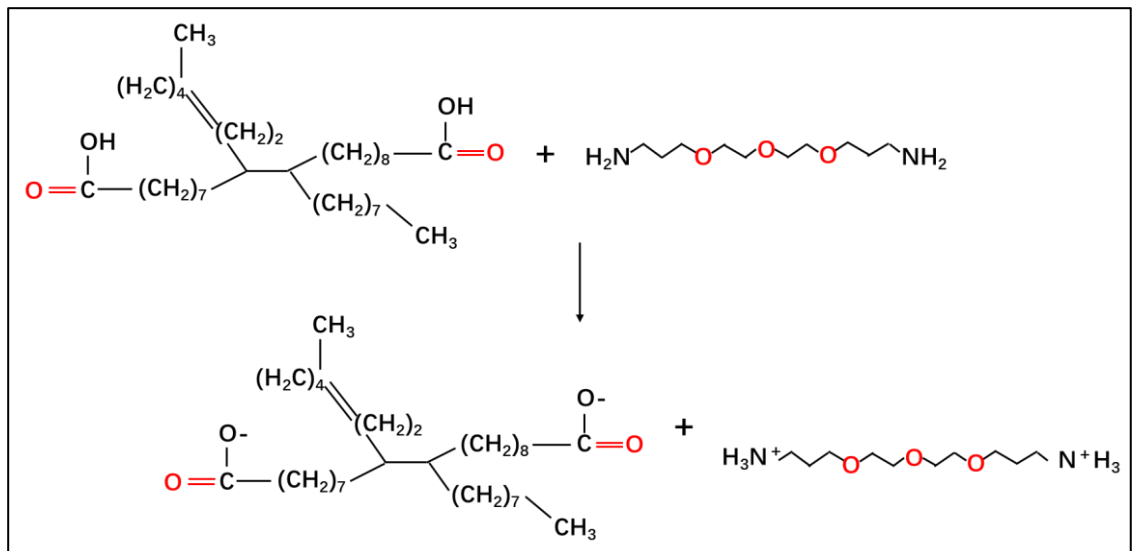

**Fig.6** The reaction to form carboxylate anions

Then, the esterification reaction is occurred between carboxylate anions and the epoxide groups in the epoxy resin, which generating the oxygen anions. Esterification reaction belongs to $S_N2$ nucleophilic substitution reaction, and the reaction rate is related to the nucleophile of epoxy group and the structure of the carboxylic acid anion [15].

The esterification reaction is described in Figure 7. For epoxy resin adhesives, the epoxy group attacks the carbon atom connected to the carboxylate anion from the backside of the O⁻ atom during the esterification reaction. It initially forms a relatively weak bond with the carbon atom, while the bond between the leaving group O⁻ and the carbon atom is also weakened to some extent. Both the epoxy group and the leaving group O⁻ are in line with the carbon atom, and the other three bonds on the carbon atom gradually transform from umbrella-like to planar. At this point, the reactants form an activated complex in the state of an intermediate transition is formed, known as the transition state. Subsequently, the bond between the epoxy group and the carbon atom begins to form, while the bond between the carbon atom and the leaving group O⁻ breaks. The three bonds on the carbon atom shift from the plane to the other side, resembling the flipping of an umbrella from inside out by a strong wind. At this stage, energy is released and the product is formed.

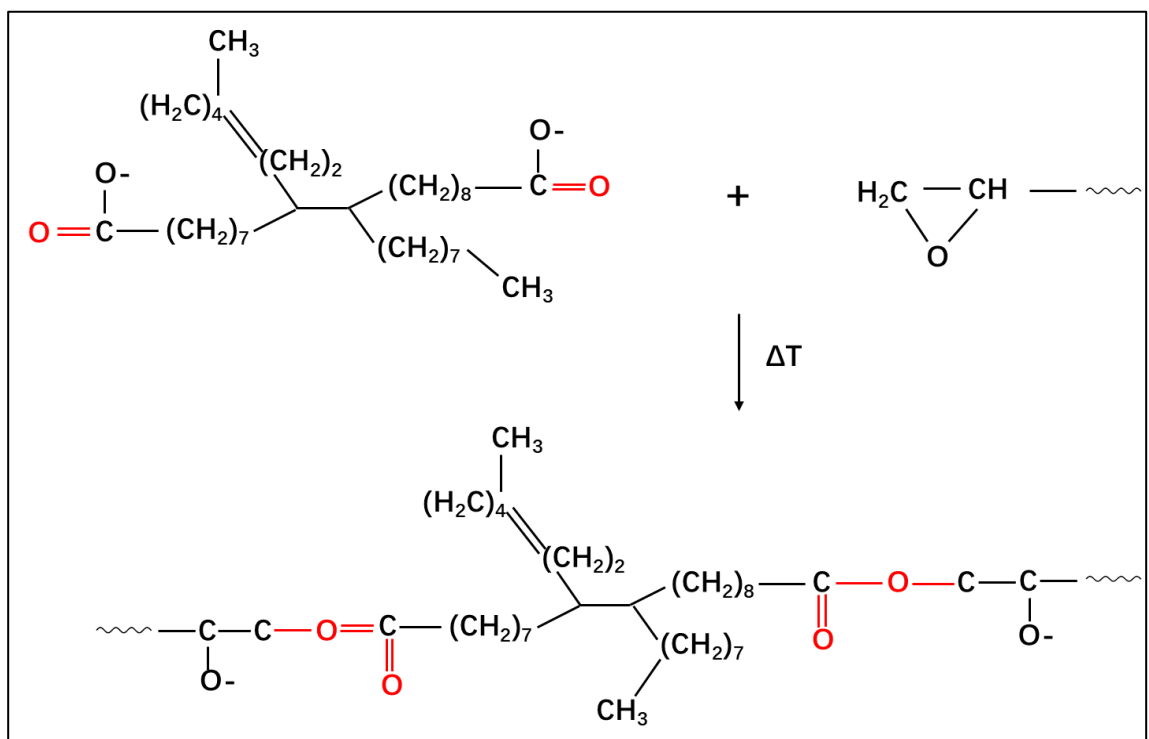

**Fig.7** The esterification reaction between carboxylate anions and the epoxide groups

The transition state theory states that the transition state is an unstable state and represents the highest point of potential energy, which is the most energetically unfavorable state to reach. This process requires the absorption of activation energy ($E_a$) [16]. The energy of the molecules is enhanced with the increasing of the temperature, which allowing more molecules to reach the activated state. The more molecules reach the transition state, the faster the reaction rate. Therefore, the esterification reaction proceeds slowly at room temperature and requires methods such as increasing the reaction temperature to enhance the reaction rate.

After the esterification, the oxygen anions react with another unsaturated fatty acid to form new carboxylate anions, namely etherification. The esterification-etherification reactions were carried out repeatedly until the crosslinking and curing of the adhesive were completed.

Usually, esterification reactions require external heating to accelerate, which is extremely slow at room temperature. It can be considered that the rate-determining step in the pre-curing reaction of the 3M epoxy adhesive at room temperature is the esterification. Therefore, the shear strength of the joints bonded with 3M epoxy adhesive can only reach 0.87 MPa after a pre-curing time of 4 hours, without external heating.

**4.3 Pre-curing kinetics analysis**

The pre-curing process of adhesives has a significant impact on the quality of bonding. There are three fundamental process parameters in curing: temperature, pressure and time. During isothermal pre-curing, the chemical reaction rate equation can be described as:

$$\frac{d\alpha}{dt} = k(T) \cdot (1-\alpha)^n$$

The Arrhenius equation corresponds to the constant k(T) and depends on the pre-exponential factor A, the activation energy $E_\alpha$ and the temperature T. In this context, the constant R means the general gas constant (8.314 J/mol·K) [17].

$$k(T) = A \cdot exp(-\frac{E_\alpha}{RT})$$

The curing reaction kinetics model is established by integrating above equation for the adhesive system:

$$\alpha(t) = 1 - [A \cdot exp\left(-\frac{E_\alpha}{RT}\right)(n-1)t + 1]^{\frac{1}{1-n}}$$

In the equation, α(t) is the degree of cure, n is the reaction order and T is the temperature. From this equation, it can be observed that the pre-curing characteristics are influenced by both temperature and curing time. Higher temperature and longer time result in a higher degree of curing. Therefore, the DSC analysis is conducted on the curing process of the two adhesive systems to determine the curing time and temperature.

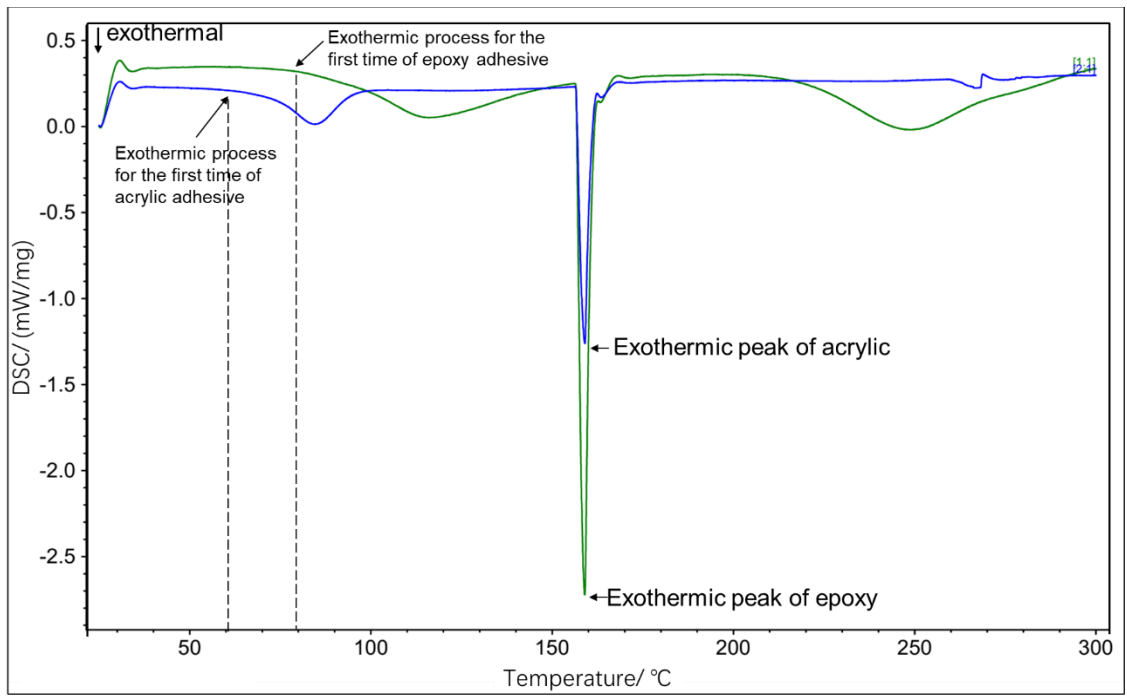

**Fig.8** The DSC curves of two adhesive systems during a heating rate of 10 °C/mi

Figure 8 shows the DSC curves of two adhesive systems during a heating rate of 10 °C/min. The area of exothermic peak corresponds to the heat released during the curing reaction. From the DSC results, it can be observed that both adhesive systems require a baking temperature of 150°C or above for complete curing. The acrylic adhesive of Lord exhibits a noticeable exothermic reaction starting at around 80°C, while the epoxy adhesive of 3M shows curing exotherm at around 60°C. As reviewed in the above paragraphs, the free-radical polymerization reaction of the two-component adhesive of acrylic system is very fast at room temperature, which achieving sufficient pre-curing strength quickly for deformation control. On the other hand, due to the lengthy esterification during pre-curing process, the epoxy resin adhesive cannot rapidly attain effective pre-cure strength without external heating conditions. Therefore, for the two-component adhesive of the epoxy resin system, it is recommended to perform a pre-cure treatment at 80°C to accelerate the pre-curing reaction, which could reducing the occurrence of deformation problems. In contrast, it does not have any specific requirements of heating pre-curing process for the two-component adhesive of acrylic system.

## 5 Conclusion

In this study, pre-curing mechanism at room temperature of epoxy and acrylate adhesive are analyzed for deformation control of automobile hood. The pre-curing strength before high-temperature baking is the key influencing factor on the adhesive deformation, and the shorter the time it takes to form a certain strength, the easier it is to control the amount of deformation.

(1) The key of the pre-curing time of acrylic adhesives at the room temperature lies in the chain initiation stage of the free radical polymerization reaction. The shear strength of this joint can reach to 3.67 MPa with a pre-curing time of 1 h, which quickly achieving the pre-cure strength required for deformation control.

(2) For epoxy resin adhesives, the rate-determining step in the pre-curing reaction process is the esterification reaction. Due to the influence of the structure of carboxylic acid and the nucleophile of epoxy group, the esterification reaction process is relatively long. The shear strength of this joint can only reach 0.87 MPa after pre-curing for 4 h without external heating.

(3) Since the inability to quickly achieve effective pre-curing strength of the epoxy resin adhesives, it is recommended to conduct heating pre-curing treatment at 80 °C to accelerate the pre-curing reaction, which could reduce the occurrence of deformation. In contrast, it does not have any specific requirements of heating pre-curing process for the two-component adhesive of acrylic system.

## Acknowledgments

Funding: This work was supported by the Scientific and Technological Research Program of Chongqing Municipal Education Commission (grant number KJQN202203205).